\documentstyle[psfig]{mn}

\newif\ifAMStwofonts

\def\HII{H{\sevensize II}}

\ifoldfss
  \ifCUPmtlplainloaded \else
    \NewTextAlphabet{textbfit} {cmbxti10} {}
    \NewTextAlphabet{textbfss} {cmssbx10} {}
    \NewMathAlphabet{mathbfit} {cmbxti10} {} 
    \NewMathAlphabet{mathbfss} {cmssbx10} {} 
  \fi
  \ifAMStwofonts
    \ifCUPmtlplainloaded \else
      \NewSymbolFont{upmath} {eurm10}
      \NewSymbolFont{AMSa} {msam10}
      \NewMathSymbol{\upi}     {0}{upmath}{19}
      \NewMathSymbol{\umu}     {0}{upmath}{16}
      \NewMathSymbol{\upartial}{0}{upmath}{40}
      \NewMathSymbol{\leqslant}{3}{AMSa}{36}
      \NewMathSymbol{\geqslant}{3}{AMSa}{3E}

      \let\leq=\leqslant 
       
    \fi
  \fi
\fi 

\ifnfssone
  \newmathalphabet{\mathit}
  \addtoversion{normal}{\mathit}{cmr}{m}{it}
  \addtoversion{bold}{\mathit}{cmr}{bx}{it}
  \newmathalphabet{\mathbfit} 
  \addtoversion{normal}{\mathbfit}{cmr}{bx}{it}
  \addtoversion{bold}{\mathbfit}{cmr}{bx}{it}
  \newmathalphabet{\mathbfss} 
  \addtoversion{normal}{\mathbfss}{cmss}{bx}{n}
  \addtoversion{bold}{\mathbfss}{cmss}{bx}{n}
  \ifAMStwofonts
    \ifCUPmtlplainloaded \else
      \UseAMStwoboldmath
      \makeatletter
      \new@mathgroup\upmath@group
      \define@mathgroup\mv@normal\upmath@group{eur}{m}{n}
      \define@mathgroup\mv@bold\upmath@group{eur}{b}{n}
      \edef\UPM{\hexnumber\upmath@group}
      \new@mathgroup\amsa@group
      \define@mathgroup\mv@normal\amsa@group{msa}{m}{n}
      \define@mathgroup\mv@bold\amsa@group{msa}{m}{n}
      \edef\AMSa{\hexnumber\amsa@group}
      \makeatother
      \mathchardef\upi="0\UPM19
      \mathchardef\umu="0\UPM16
      \mathchardef\upartial="0\UPM40
      \mathchardef\leqslant="3\AMSa36
      \mathchardef\geqslant="3\AMSa3E

      \let\leq=\leqslant 

    \fi
  \fi
\fi 

\ifnfsstwo
  \DeclareMathAlphabet{\mathbfit}{OT1}{cmr}{bx}{it}
  \SetMathAlphabet\mathbfit{bold}{OT1}{cmr}{bx}{it}
  \DeclareMathAlphabet{\mathbfss}{OT1}{cmss}{bx}{n}
  \SetMathAlphabet\mathbfss{bold}{OT1}{cmss}{bx}{n}
  \ifAMStwofonts
    \ifCUPmtlplainloaded \else
      \DeclareSymbolFont{UPM}{U}{eur}{m}{n}
      \SetSymbolFont{UPM}{bold}{U}{eur}{b}{n}
      \DeclareSymbolFont{AMSa}{U}{msa}{m}{n}
      \DeclareMathSymbol{\upi}{0}{UPM}{"19}
      \DeclareMathSymbol{\umu}{0}{UPM}{"16}
      \DeclareMathSymbol{\upartial}{0}{UPM}{"40}
      \DeclareMathSymbol{\leqslant}{3}{AMSa}{"36}
      \DeclareMathSymbol{\geqslant}{3}{AMSa}{"3E}

      \let\leq=\leqslant 

    \fi
  \fi
\fi 

\ifCUPmtlplainloaded \else
  \ifAMStwofonts \else 
    \def\upi{\pi}
    \def\umu{\mu}
    \def\upartial{\partial}
  \fi
\fi

\title{Chemical abundances and ionizing clusters of HII regions in the LINER galaxy NGC 4258}
\author[A. I. D\'\i az et al.]
       {Angeles I.~D\'\i az,$^1$ Marcelo~Castellanos,$^1$ Elena~Terlevich$^2$ 
\thanks{Visiting Fellow, IoA, Cambridge} and
\newauthor
Mar\'\i a Luisa~Garc\'\i a-Vargas$^3$\\
        $^1$Departamento de F\'\i sica Te\'orica, C-XI, Universidad Aut\'onoma de Madrid, 28049 Madrid, Spain\\
        $^2$INAOE, Tonantzintla, Apdo. Postal 51, 72000 Puebla, M\'exico\\
        $^3$Grantecan S.A., 38200 La Laguna, Tenerife, Spain}
\date{Accepted 
      Received ;
      in original form }

\pagerange{\pageref{firstpage}--\pageref{lastpage}}
\pubyear{1999}

\begin{document}

\maketitle

\label{firstpage}

\begin{abstract}

We present long-slit observations in the optical and near infrared
of eight H {\sevensize II} regions in the spiral galaxy NGC 4258. Six of the
observed regions are located in the SE inner spiral arms and the other two are 
isolated in the northern outer arms. 

A detailed analysis of the physical conditions of the gas has been performed. 
For two of the regions an electron temperature has been derived from the 
[SIII] $\lambda$ 6312 \AA\ line. For the rest, an empirical calibration based 
on the red and near infrared sulphur lines has been used. The oxygen 
abundances derived by both methods are found to be significantly lower (by a 
factor of two) than previously derived by using empirical calibrations based 
on the optical oxygen lines. 

In the brightest region, 74C, the observation of a prominent feature due to 
Wolf-Rayet stars provides an excellent constraint over some properties of 
the ionizing clusters. In the light of the current evolutionary synthesis 
models, no consistent solution is found to explain at the same time both the 
WR feature characteristics and the emission line spectrum of this region. In 
principle, the presence of WR stars could lead to large temperature 
fluctuations and also to a hardening of the ionizing radiation. None of these 
effects are found in region 74C for which the electron temperatures found from 
the [SIII] $\lambda$ 6312 \AA\ line and the Paschen discontinuity at 8200 
\AA\ are equal within the errors and the effective temperature of the ionizing 
radiation is estimated at around 35300 K. 

Both more observations of confirmed high metallicity regions and a finer 
metallicity grid for the evolutionary synthesis models are needed in order to 
understand the ionizing populations of HII regions.

\end{abstract}


\section{Introduction}

Theoretically, the evolution of a young stellar population depends on 
metallicity at least 
through two very important effects: the increasing opacity of the stellar 
material and the dependence of mass loss on metal content in high mass stars.
As a consequence of the first, the effective temperature of ionizing stars 
should be  
lower in regions of higher metallicity (see for example McGaugh 1991) but, on 
the other hand, as a consequence of the second, that is if the strength
of stellar winds increases with metallicity, the loss of the outer
envelopes of the most massive stars can increase their surface temperature 
to very high values. These highly evolved massive O stars are identified with 
the Wolf Rayet population. 

Most theoretical evolutionary models for ionizing star clusters predict the 
appearance of WR stars -- although at slightly different cluster ages 
depending on the specific details of the models -- and in all of them the 
fraction of WR stars increases with metallicity since the limiting mass for a 
star to enter the WR phase decreases with increasing metallicity. However, the 
effect of these stars on their surrounding ionized gas depends on the 
characteristics of the assumed wind opacity and therefore while some models 
predict the existence of high metallicity H{\sevensize II} regions of high 
excitation (e.g. Garc\'\i a-Vargas \& D\'\i az 1994; Stasi\'nska \& 
Leitherer 1996), others do not (e.g. Cervi\~no \& Mas-Hesse 1994). 
The finding of this extreme HII region population would constitute 
a firm evidence of the existence of WR stars with high effective temperatures, 
thus favoring the first kind of models but, even if these regions are not 
definitively identified, the study of high metallicity H{\sevensize II} regions
is of crucial importance to investigate the processes of star formation and 
evolution in high metallicity environments. This investigation gains even more
relevance if we take into account the recent observations that point out to a 
firm connection between star formation and activity in active galactic nuclei
(Heckman et al. 1997, Gonz\'alez-Delgado et al. 1998) already predicted on 
theoretical grounds (Terlevich \& Melnick 1985; Terlevich et al. 1992), since 
these nuclei seem to reach rather high metallicities (Phillips et al. 1984;
Gonz\'alez-Delgado \& P\'erez 1996) as deduced from the study of their 
circumnuclear star forming regions.
  
Unfortunately, the term ``high metallicity H{\sevensize II} region" is rather 
ill defined. Most authors apply the term to H{\sevensize II} regions of solar 
or oversolar metal content, but the difficulty to derive directly the 
metallicity of these cool HII regions is well known and the empirical 
calibrations commonly employed provide estimations with an uncertainty larger 
than a factor of 2 (see e.g. 
D\'\i az et al. 1991). On the other hand, the characteristics of WR features 
are highly dependent on metallicity. Both the luminosity of the WR ``bump", 
that 
is the combination of the N{\sevensize III} and the He{\sevensize II} lines at 
around $\lambda$ 4660 \AA , and its equivalent width are predicted to be 
largest at the 
highest metallicity and almost an order of magnitude difference is found 
between the computed WR ``bump" intensity relative to H$\beta$ in clusters of 
half solar and 
solar metallicities (Schaerer \& Vacca 1998). Therefore, the detection of WR 
features combined with a detailed analysis of the emission line spectra 
of suspected high metallicity H{\sevensize II} regions, can constitute 
excellent tools for the simultaneous determination of both the metallicity and 
the age rather accurately in these regions. 

In the frame of a long term programme to study high metallicity H{\sevensize 
II} regions we are presently carrying out a search for the high metallicity 
and high excitation  H{\sevensize II} region population by performing 
spectrophotometric observations at a resolution high enough to detect and 
measure WR features. For that, we have selected from the literature
the giant HII regions with 
solar or oversolar abundance, as deduced from empirical calibrations based on 
the optical forbidden lines, showing the highest excitation. This corresponds 
to a ratio of [OIII]/H$\beta$ $\sim$ 1.0, which is actually quite moderate
but still high compared to that expected in high metallicity HII regions.
Here we present the first of these investigations involving several \HII\ 
regions in the galaxy NGC~4258.

NGC~4258 (M106) is classified as SAB(s)bc and Sb(s)II types by De 
Vaucouleurs \shortcite{deva} and Sandage \& Tammann \shortcite{santam} 
respectively. Heckman \shortcite{hec} and Stauffer \shortcite{stau} 
classified its nucleus as belonging to a transition type between LINER
and H {\sevensize II}-region like emission-line galaxy. As it is addressed by 
Court\'{e}s et al. \shortcite{cou}(hereinafter C93), the H {\sevensize II}
region distribution is very peculiar in the sense that three general structures
are found: the normal spiral arms, with dense and very bright H{\sevensize II} 
regions following pure spiral shapes as a consequence of the barred spiral 
type of this galaxy; the faint outer arms (H {\sevensize II} region poor), and 
the anomalous spiral arms, discovered by Court\'{e}s \& Cruvellier 
\shortcite{coucru}, without star formation and with no evidence of very blue 
stars \cite{depe}. It is assumed that the anomalous arms 
were formed by powerful jets ejected by the active nucleus of the galaxy. 
St\"{u}we, Schulz \& H\"{u}hnermann \shortcite{SSH} showed that the active 
nucleus must be obscured along the line of sight as is also indicated by its 
invisibility in IUE spectra. Cecil, Morse \& Veilleux \shortcite{CMV} found, 
from the emission-line flux ratios, LINER-like gaseous excitation along the 
jets. They also point out the possibility that the circumnuclear region has 
undergone recent formation of massive stars induced by the jets that are 
flowing into the disk. However, no evidence for this star formation has been 
found, maybe due to the obscuration by molecular gas along the line of sight. 
Finally, a subparsec diameter gaseous disk shows keplerian motion around the
centre of the galaxy, which indicates the presence of  a mass of 3.6 $\times$ 
10$^{7}$ M$_{\odot}$ within a radius of less than 0.13 pc \cite{miy}. This is 
probably the best existing evidence for the presence of a massive black hole 
at the centre of a galaxy.

The problem of the distance determination for this galaxy still remains 
unsolved. The quoted distance values range from 3.4 Mpc \cite{RAJ} to 10.4 Mpc 
\cite{santam}. The adopted value in the present work is 5.5 Mpc \cite{mar}.

We have observed eight H{\sevensize II} regions along two different spiral 
arms. Four of them have been identified as:  74C, 69C, 5N(A) and 5N(B) in C93. 
Regions 5N(A) and 5N(B) are located in the northern outer arm at a 
galactocentric distance of 20.3 Kpc. The other six regions are located in the 
SE inner arm at 8.05 Kpc from the galaxy centre. Region 74C is a discrete 
radiosource at 4.9 GHz \cite{hum}.

Regions 74C and 5N were previously observed by Oey \& Kennicut \shortcite{ok} 
who derived  values of 12+log(O/H) of 8.87 and 8.74 for each of them 
respectively (the solar value is 8.92). 
With a ratio of [O{\sevensize III}] to H$\beta$ of 0.98 and 1.44 both regions
fit our selection criteria.

\section[]{Observations and Reductions}

Our spectrophotometric observations were obtained with the 4.2m William 
Herschel Telescope at the Roque de los Muchachos Observatory, in 1994 March 13,
using the ISIS double spectrograph, with the TEK1 and a EEV CCD detectors in 
the blue and red arm respectively. Two gratings were used, R300B in the blue 
and R316R in the red arm, covering from [O {\sevensize II}] $\lambda$3727 to 
[S {\sevensize III}] $\lambda$9532 in three different spectral ranges of 
$\sim$1700 {\AA} each. The dispersion of 1.4 {\AA} pixel$^{-1}$ with a slit 
width of 1\farcs5 gives a spectral resolution of $\sim$4 {\AA}. The nominal 
spatial sampling is 0\farcs7 pixel$^{-1}$. The seeing was $\sim$1\farcs5. 
The slit was centered at the positions of regions 74C and 5N, as given by 
Oey \& Kennicutt (1993). A third position was observed in order to include 
other HII regions outlining the SE inner arm. In all cases the position angle was chosen as to maximize the number of HII rgions in the slit.
A journal of observations is 
given in Table 1.

The data were reduced using the IRAF (Image Reduction and Analysis Facility)
package following standard methods. The
two-dimensional wavelength calibration was accurate to 1 {\AA} in all cases.
The two-dimensional frames were flux calibrated using two spectroscopic 
standard stars observed
during the same night with a 5\farcs\ width slit. The agreement between the
individual calibration curves was better than 8\% in all cases. The spectra were previously corrected for atmospheric extinction using a mean extinction curve applicable to La Palma observing site. 
The removal of the atmospheric water-vapour absorption bands in the near
infrared \cite{dipawi}, was achieved dividing by the relatively featureless continuum of a subdwarf star observed in the same night as the galaxy.


\begin{figure}
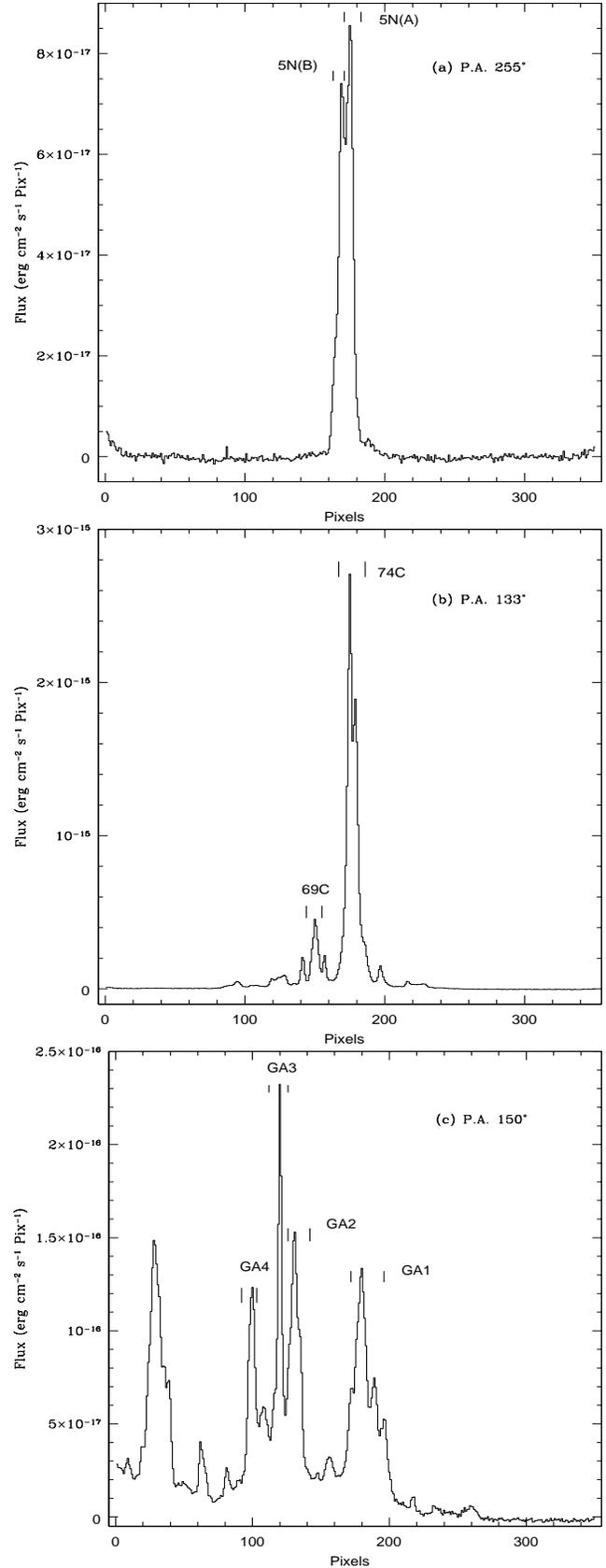

 \psfig{figure=pos5N.epsi,width=8.4cm,height=7.4cm,clip=}
 \psfig{figure=posWR.epsi,width=8.4cm,height=7.4cm,clip=}
 \psfig{figure=posGA.epsi,width=8.4cm,height=7.4cm,clip=}
\caption{H$\alpha$ profiles for the three observed slit positions including:
(a) regions 5N(A) and 5N(B); (b) regions 74C and 69C; (c) 
regions GA1, GA2, GA3 and GA4.}
\end{figure}


The most critical step in the reduction process was the background 
subtraction in the near infrared frames. There are numerous OH emission lines
of the night-sky spectrum that contaminate the near infrared spectra. This
contamination is particularly important in the range from
9376 to 9600 {\AA} \cite{osfubi}. Hence, the [S{\sevensize III}] $\lambda$9532 {\AA} and Pa 8 lines intensities could be affected even after the background
subtraction is made. Fortunately, the [S{\sevensize III}] $\lambda$9069 {\AA} 
and Pa 9 lines are unaffected by these night-sky lines. The observed 
$\lambda$9532/$\lambda$9069 ratio is above the theoretical value of 2.44 in
four of the observed regions. In these cases, the [S{\sevensize III}] 
$\lambda$9532 line was scaled by means of the more accurate [S{\sevensize III}]
$\lambda$9069 line intensity measurement.

\begin{table}
 \begin{minipage}{70mm}
 \caption{Journal of observations}
 \begin{tabular}{@{}lccccc@{}}
 {P.A. (\degr)\footnote{All the observations made on 1994 March 13/14}} & Grating & $\lambda$ range ({\AA}) & Exposure (s) \\
 255 & R300B & 3675-5300 & 1800   \\
 255 & R300B & 3675-5300 & 1800   \\
 255 & R316R & 5890-7600 & 1200   \\
 255 & R316R & 5890-7600 & 1800   \\
 255 & R316R & 7960-9670 & 1800   \\
 255 & R316R & 7960-9670 & 1800   \\
 150 & R300B & 3675-5300 & 1800   \\
 150 & R300B & 3675-5300 & 1800   \\
 150 & R300B & 3675-5300 & 1200   \\
 150 & R316R & 5890-7600 & 1800   \\
 150 & R316R & 7960-9670 & 1800   \\
 150 & R316R & 7960-9670 & 1800   \\
 133 & R300B & 3675-5300 & 1800   \\
 133 & R300B & 3675-5300 & 1200   \\
 133 & R316R & 5890-7600 & 1800   \\
 133 & R316R & 7960-9670 & 1800   
 \end{tabular}
 \end{minipage}
\end{table}


\section[]{Results}

Figure 1 shows the spatial distribution of the H$\alpha$ flux along the slit 
for the three different positions observed. Regions 5N A and B are clearly
identified on the H$\alpha$ profile corresponding to PA=255\degr . Also regions 
74C and 69C, showing a certain degree of substructure, are resolved at position
angle PA=133\degr . Four different regions are resolved at PA=150\degr that we 
have named GA1, GA2, GA3 and GA4. A fifth region is clearly seen in the 
H$\alpha$ profile that is not detected in the near IR frames.

The spectra corresponding to each of the identified regions is shown in Figs. 2 
to 4.
WR features around $\lambda$ 4680 \AA\ are seen in the spectrum of region 74C 
and to a lesser extent in that of region 5N.


\begin{figure*}
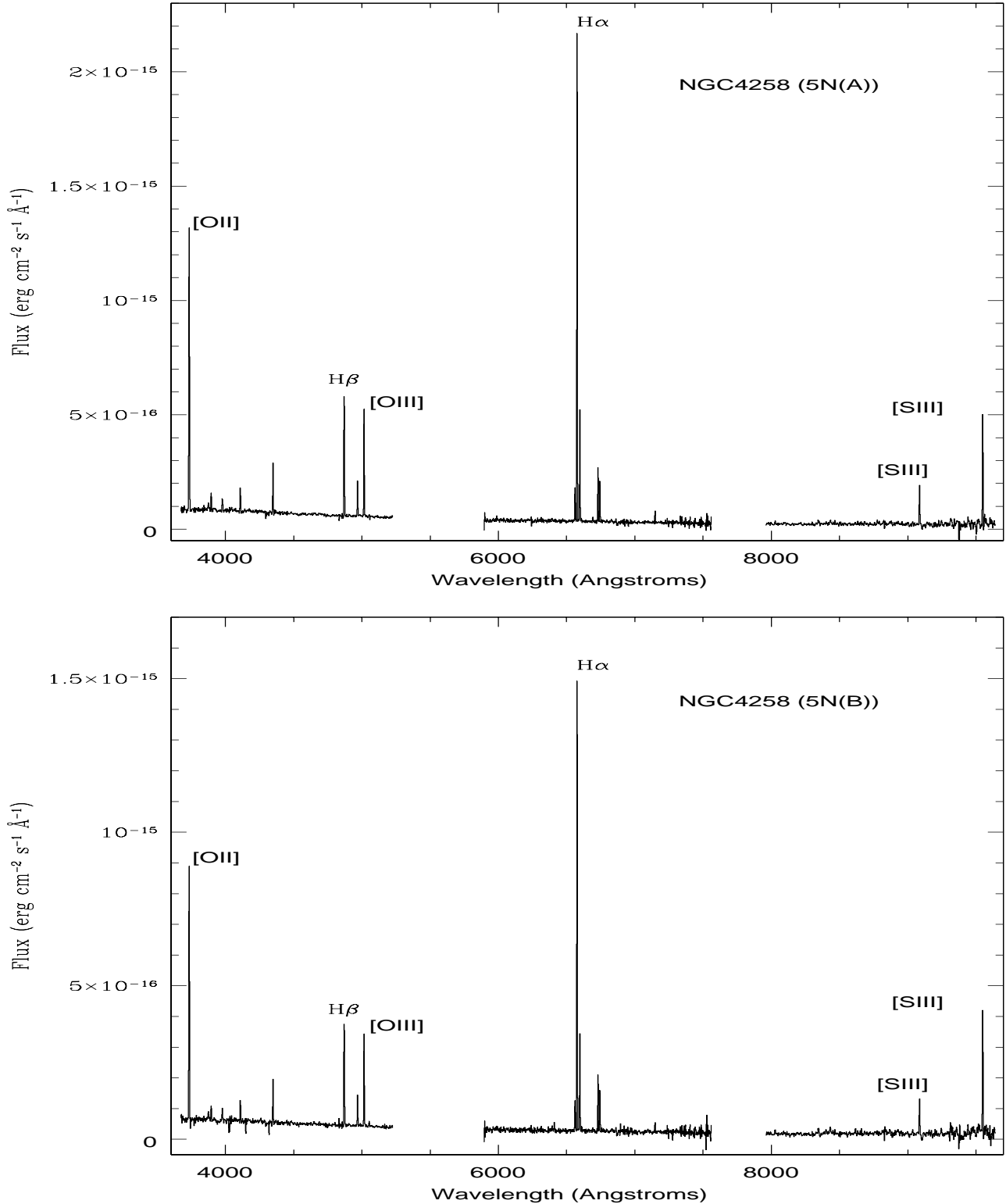

\begin{minipage}{180mm}
 \psfig{figure=5NA.epsi,height=10cm,width=17cm,clip=}
\vspace{12pt}
 \psfig{figure=5NB.epsi,height=10cm,width=17cm,clip=}
\vspace{12pt}
\caption{Merged spectra for the regions 5N(A) and 5N(B)}
\end{minipage}
\end{figure*} 

\begin{figure*}
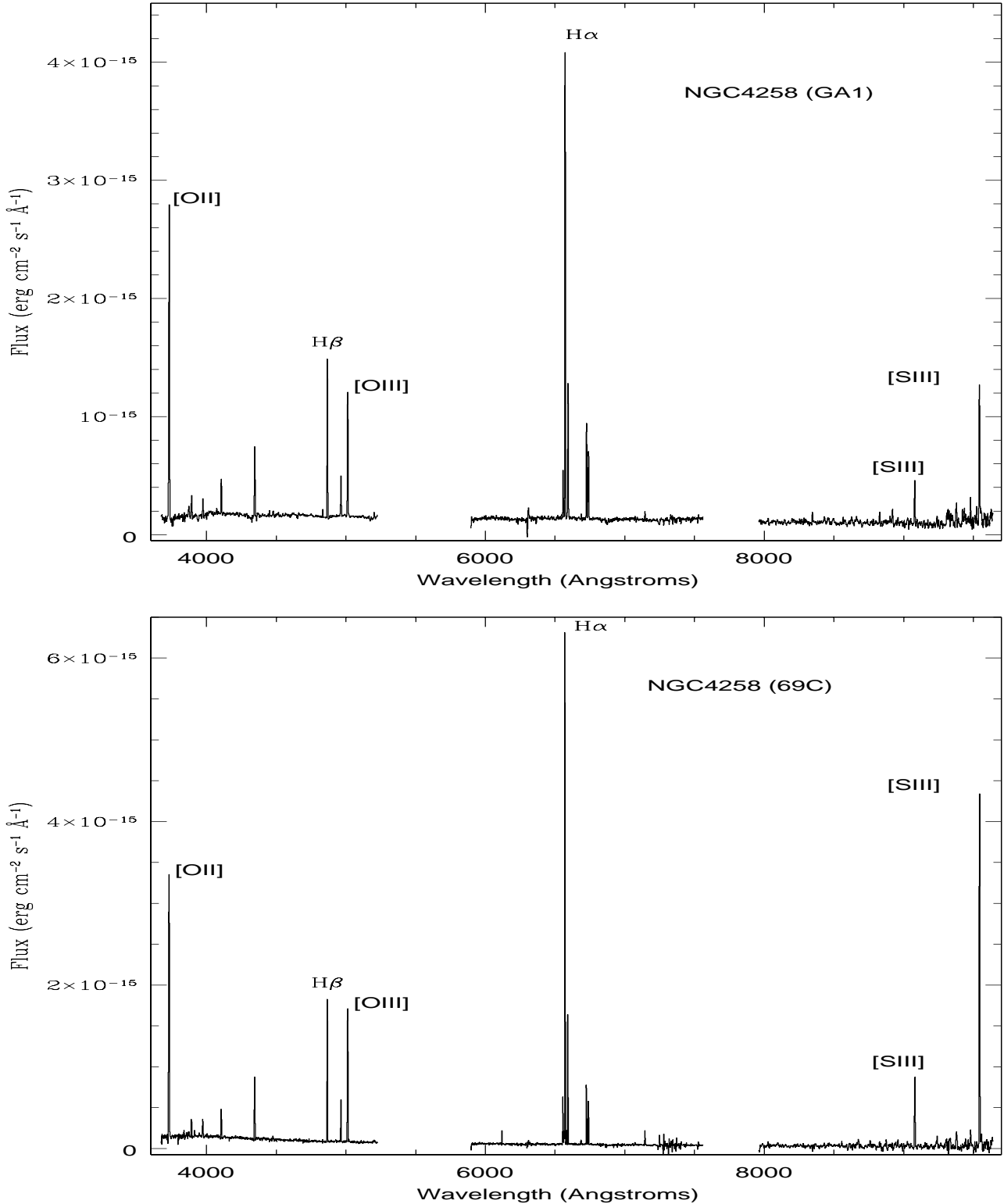

\begin{minipage}{180mm}
\psfig{figure=GA1.epsi,height=10cm,width=17cm,clip=}
\vspace{12pt}
 \psfig{figure=69C.epsi,height=10cm,width=17cm,clip=}
\vspace{12pt}
\caption{Merged spectra for the regions GA1 and 69C. They could correspond to
different regions of the same nebula}
\end{minipage}
\end{figure*}

\begin{figure*}
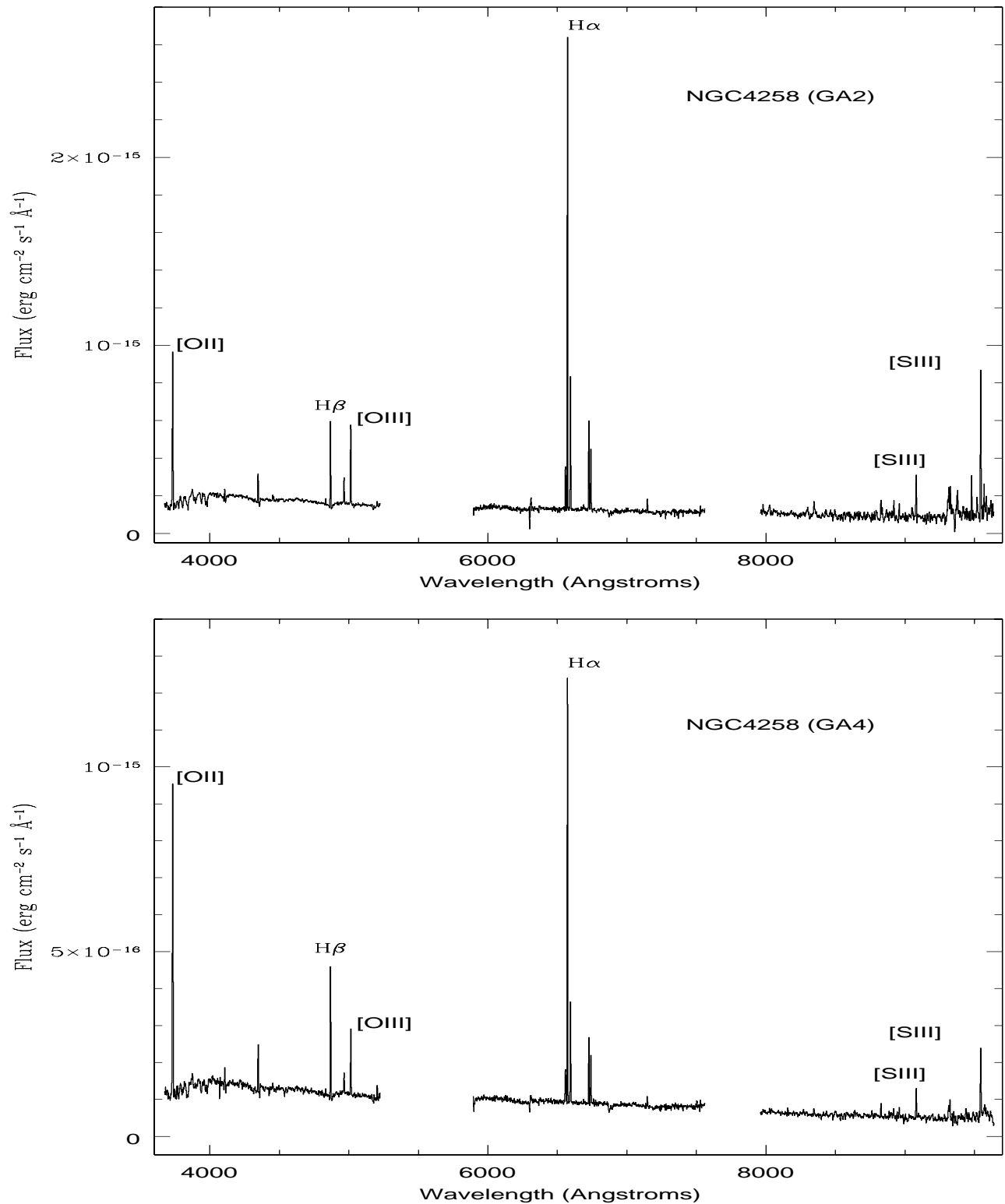
 
\begin{minipage}{180mm}
\psfig{figure=GA2.epsi,height=10cm,width=17cm,clip=}
\vspace{12pt}
\psfig{figure=GA4.epsi,height=10cm,width=17cm,clip=} 
\vspace{12pt}
\caption{Merged spectra for the regions GA2 and GA4}
\end{minipage}
\end{figure*}

\begin{figure*}
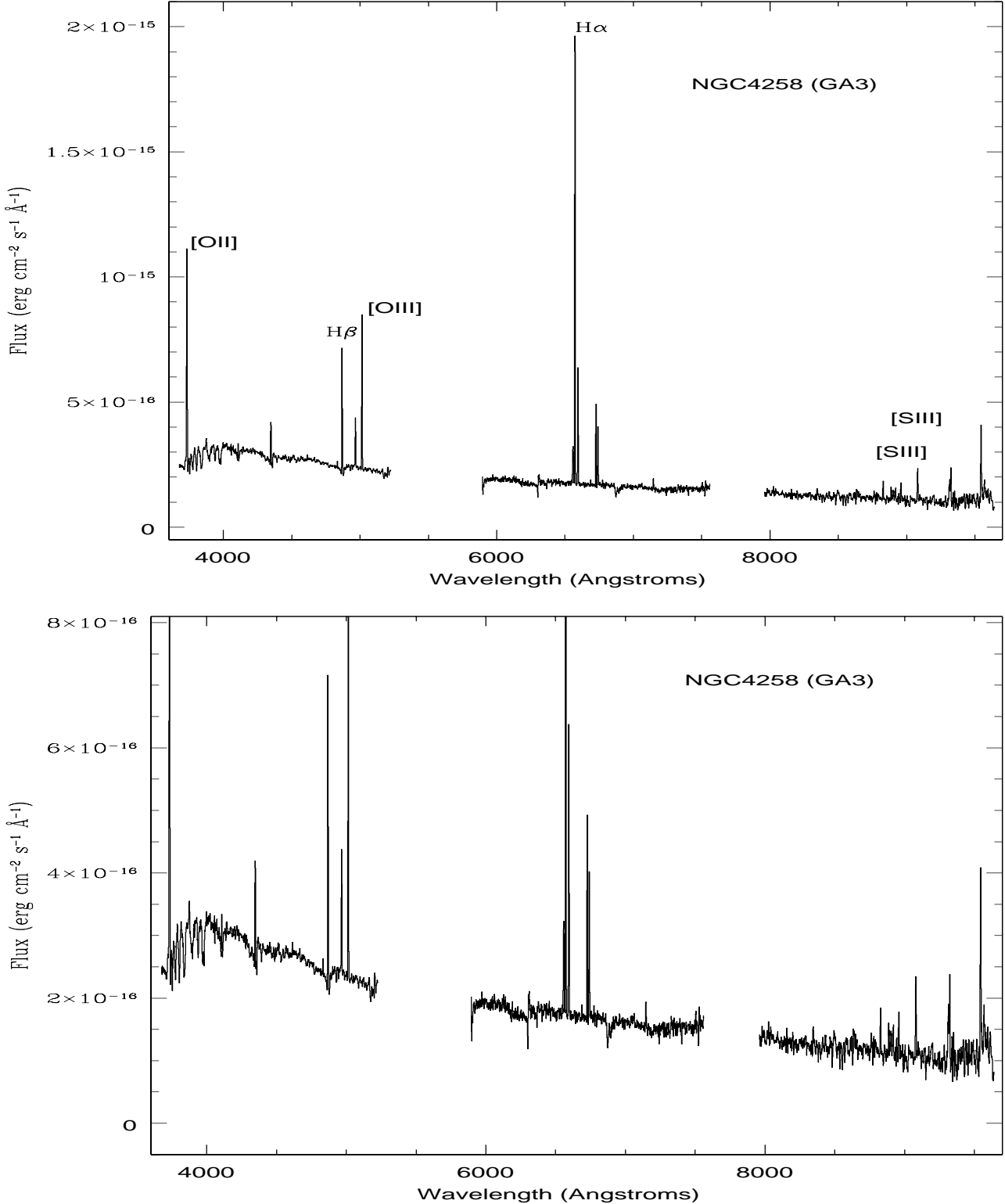

\begin{minipage}{180mm}
 \psfig{figure=GA3.epsi,height=10cm,width=17cm,clip=}
\vspace{12pt}
 \psfig{figure=scGA3.epsi,height=10cm,width=17cm,clip=}   
   \caption{Merged spectra for region GA3. The bottom panel is an enlargement 
to show the prominent absorption features.}
\end{minipage}
\end{figure*} 

\begin{figure*}
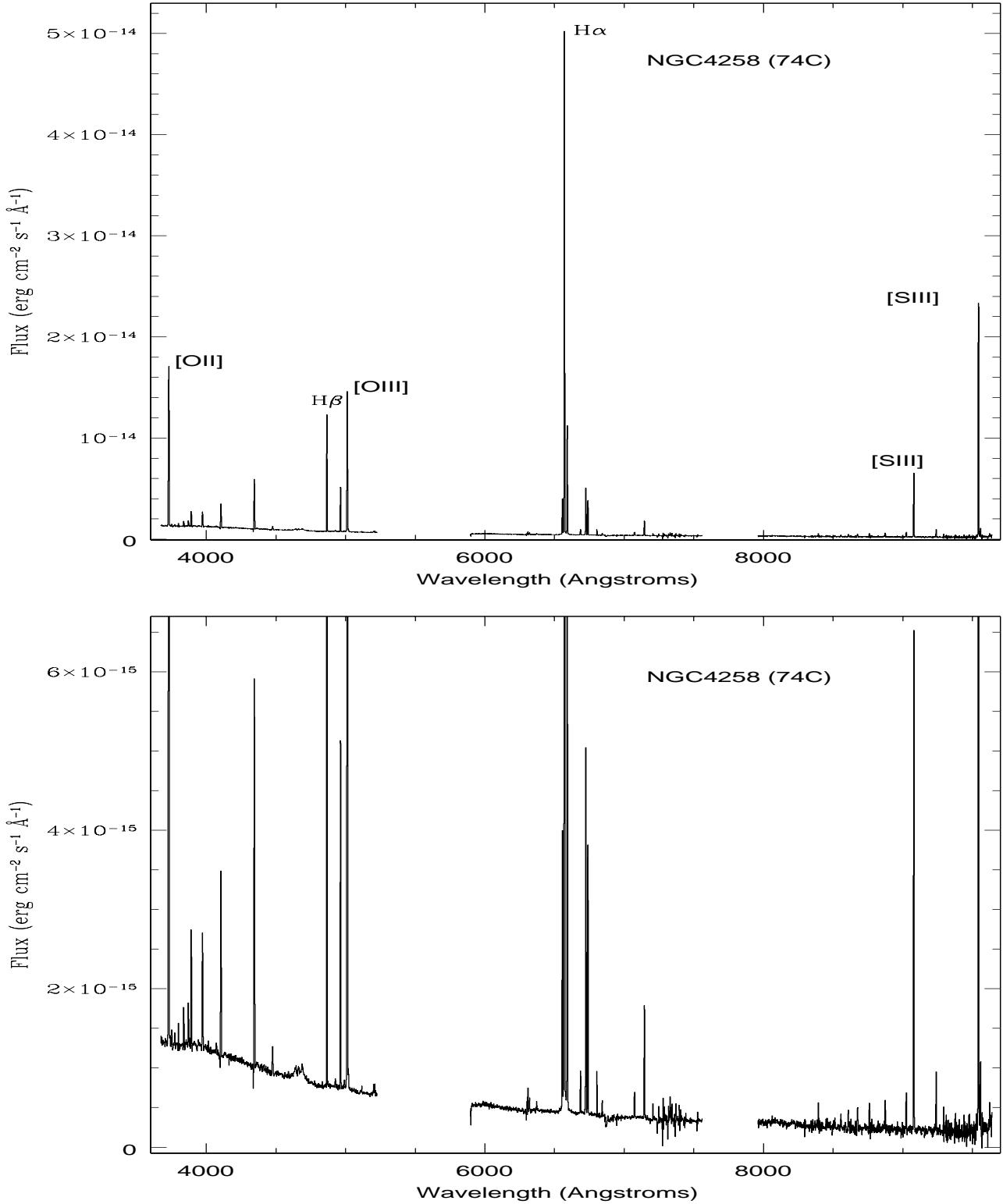

\begin{minipage}{180mm}
 \psfig{figure=74C.epsi,height=10cm,width=17cm,clip=}
\vspace{12pt}
 \psfig{figure=sc74C.epsi,height=10cm,width=17cm,clip=}   
 \caption{Merged spectra with two intensity scales for region 74C. Notice the 
WR feature at 4686 {\AA}}
\end{minipage}
\end{figure*}


\subsection{Line intensities}

Emission line fluxes were measured using the IRAF SPLOT software package, by 
integrating the line intensity over a local fitted continuum. The errors in 
the line fluxes have been calculated from the expression
$\sigma_{l}$ = $\sigma_{c}$N$^{1/2}$[1 + EW/(N$\Delta$)]$^{1/2}$, where 
$\sigma_{l}$ is the error in the line flux, $\sigma_{c}$ represents the 
standard deviation in a box near the measured emission line and stands for the 
error in the continuum placement, N is the number of pixels used in the 
measurement of the line flux, EW is the line equivalent width, and
$\Delta$ is the wavelength dispersion in angstroms per pixel.

The observed line intensities relative to the H$\beta$ line were corrected for 
interstellar reddening according to an average extinction curve \cite{ost} and 
assuming the Balmer line theoretical values for  case B recombination
\cite{brock}.
The presence of an underlying stellar population is evident in the blue 
spectra of regions GA2, GA3 and GA4. The H$\gamma$ and H$\delta$ Balmer lines 
are clearly affected by this stellar absorption. An iterative process was 
applied in order to fit observed and theoretical Balmer line intensities and 
to obtain the reddening constant c(H$\beta$). In all cases, H$\alpha$, 
H$\gamma$ and H$\delta$ were fitted, except in regions GA3 and GA4 for which 
H$\delta$ seems to be  underestimated. Reddening corrected Paschen
lines, when measured, are consistent with their theoretical values.

 The reddening
corrected line intensities together with their corresponding errors are given 
in Table 2 for regions 74C and 69C and Table 3 for the rest of the regions. 
Also given in the tables are the extinction corrected H$\alpha$ flux and the H$\beta$  
equivalent width. For all the regions, H$\alpha$ luminosities are lower than 10$^{38}$ erg s$^{-1}$, with the exception of region 74C, in which a value of 10$^{39.3}$ is found. According to Kennicutt \shortcite{kenn}, this region can be classified as a supergiant H{\sevensize II} region.


\begin{table*}
\setcounter{table}{1}
 \begin{minipage}{100mm}
  \caption{Reddening corrected line intensities for P.A.= 133\degr}
  \begin{tabular}{llcc}
  Region & & 74C & 69C \\
     ~    & &     &     \\ 
   Line   & &     &     \\
 3727 & [OII]               & 2100 $\pm$ 40 & 2280 $\pm$ 90 \\
 3751 &  H12                &   18 $\pm$  1 &  --           \\
 3770 &  H11                &   18 $\pm$  1 &  --           \\
 3798 &  H10                &   24 $\pm$  1 &  --           \\
 3835 &  H9+HeII            &   44 $\pm$  2 &   26 $\pm$  2 \\
 3868 & [NeIII]             &   53 $\pm$  2 &  --           \\
 3888 &  H8+HeI             &  161 $\pm$  3 &  118 $\pm$  7 \\
 3902 & [NeIII]+H$\epsilon$ &  151 $\pm$  4 &  122 $\pm$  6 \\
 4102 &  H$\delta$          &  251 $\pm$  5 &  220 $\pm$ 10 \\
 4340 &  H$\gamma$          &  468 $\pm$  7 &  450 $\pm$ 20 \\
 4471 &  HeI                &   33 $\pm$  2 &   30 $\pm$  2 \\
 4861 &  H$\beta$           & 1000 $\pm$  8 & 1000 $\pm$ 20 \\ 
 4922 &  HeI                &    7 $\pm$  1 &  --           \\
 4959 & [OIII]              &  378 $\pm$  5 &  290 $\pm$ 10 \\
 4987 & [FeIII]             &    7 $\pm$  1 &  --           \\
 5007 & [OIII]              & 1140 $\pm$ 10 &  930 $\pm$ 30 \\
 5012 &  HeI                &   17 $\pm$  2 &  --           \\
 5200 & [NI]                &    8 $\pm$  1 &   13 $\pm$  1 \\
 6300 & [OI]                &   17 $\pm$  1 &   20 $\pm$  1 \\
 6312 & [SIII]              &    9 $\pm$  1 &    7 $\pm$  1 \\
 6360 & [OI]                &    4 $\pm$  1 &   10 $\pm$  2 \\
 6548 & [NII]               &  200 $\pm$ 10 &  260 $\pm$ 10 \\
 6563 &  H$\alpha$          & 2860 $\pm$ 60 & 2850 $\pm$ 40 \\
 6584 & [NII]               &  610 $\pm$ 20 &  780 $\pm$ 20 \\
 6678 &  HeI                &   30 $\pm$  2 &   22 $\pm$  2 \\
 6717 & [SII]               &  260 $\pm$ 10 &  363 $\pm$  8 \\
 6731 & [SII]               &  188 $\pm$  3 &  253 $\pm$  7 \\
 7135 & [ArIII]             &   74 $\pm$  2 &   66 $\pm$  4 \\
 8502 &  P16                &    4 $\pm$  1 &  --           \\
 8545 &  P15                &    6 $\pm$  1 &  --           \\
 8598 &  P14                &    6 $\pm$  1 &  --           \\
 8665 &  P13                &   10 $\pm$  1 &  --           \\
 8750 &  P12                &   11 $\pm$  1 &  --           \\
 8863 &  P11                &   15 $\pm$  1 &  --           \\
 9014 &  P10                &   19 $\pm$  1 &  --           \\
 9069 & [SIII]              &  266 $\pm$  4 &  190 $\pm$ 10 \\
 9229 &  P9                 &   27 $\pm$  2 &   26 $\pm$  3 \\
 9532 & [SIII]              &  660 $\pm$ 20 & {930\footnote{Affected by a cosmic ray}} $\pm$ 30 \\ 
 9546 &  P8                 &   35 $\pm$  1 &   34 $\pm$  2 \\ 
 c(H$\beta$) &              &    0.48       &    0.08       \\
 {F(H$\alpha$)
 \footnote{10$^{-16}$ erg cm$^{-2}$ s$^{-1}$, corrected for reddening}}& & 5576 & 321 \\
 EW(H$\beta$)({\AA})& & 81 & 113 
\end{tabular}
\end{minipage}
\end{table*}



\begin{table*}
 \begin{minipage}{200mm}
  \caption{Reddening corrected line fluxes for P.A.= 255\degr and 150\degr}
  \begin{tabular}{@{}llcccccc@{}}
   Region & & 5N(A) & 5N(B) & GA1 & GA2 & GA3 & GA4 \\
   Line   & &       &       &     &     &     &     \\ 
 3727 & [OII] & 3300 $\pm$ 100 & 3400 $\pm$ 200 & 2340 $\pm$ 80 & 3010 $\pm$ 90 & 2100 $\pm$ 100 & 2800 $\pm$ 200 \\
 3770 &  H11 & -- & -- & 37 $\pm$ 2 & -- & -- & -- \\
 3798 &  H10 & -- & -- & 31 $\pm$ 3 & -- & -- & -- \\
 3888 &  H8+HeI & 145 $\pm$ 8 & 160 $\pm$ 10 & 133 $\pm$ 7 & -- & -- & -- \\
 3902 & [NeIII]+H$\epsilon$ & 125 $\pm$ 6 & 136 $\pm$ 9 & 105 $\pm$ 7 & -- & -- & --  \\
 4102 &  H$\delta$ & 240 $\pm$ 10 & 250 $\pm$ 20 & 240 $\pm$ 10 & 220 $\pm$ 10 & 150 $\pm$ 20 & 190 $\pm$ 30  \\
 4340 &  H$\gamma$ & 450 $\pm$ 20 & 480 $\pm$ 30 & 450 $\pm$ 10 & 400 $\pm$ 10 & 390 $\pm$ 30 & 380 $\pm$ 20  \\
 4861 &  H$\beta$ & 1000 $\pm$ 30 & 1000 $\pm$ 40 & 1000 $\pm$ 20 & 1000 $\pm$ 20 & 1000 $\pm$ 40 & 1000 $\pm$ 30  \\
 4959 & [OIII] & 290 $\pm$ 10 & 280 $\pm$ 20 & 270 $\pm$ 10 & 260 $\pm$ 10 & 390 $\pm$ 20 & 160 $\pm$ 10  \\
 5007 & [OIII] & 840 $\pm$ 30 & 840 $\pm$ 40 & 790 $\pm$ 20 & 860 $\pm$ 20 & 1140 $\pm$ 60 & 520 $\pm$ 30  \\
 5200 & [NI] & -- & -- & -- & 56 $\pm$ 2 & 38 $\pm$ 6 & -- \\
 6548 & [NII] & 190 $\pm$ 10 & 200 $\pm$ 10 & 250 $\pm$ 10 & 260 $\pm$ 10 & 240 $\pm$ 20 & 240 $\pm$ 10 \\
 6563 &  H$\alpha$ & 2820 $\pm$ 90 & 2880 $\pm$ 90 & 2800 $\pm$ 50 & 2800 $\pm$ 30 & 2800 $\pm$ 90 & 2850 $\pm$ 90 \\
 6584 & [NII] & 640 $\pm$ 20 & 630 $\pm$ 30 & 780 $\pm$ 20 & 790 $\pm$ 10 & 760 $\pm$ 30 & 700 $\pm$ 20 \\
 6678 &  HeI & 28 $\pm$ 2 &  & 22 $\pm$ 2 & 21 $\pm$ 2 & 17 $\pm$ 3 & 15 $\pm$ 2   \\
 6717 & [SII] & 310 $\pm$ 20 & 350 $\pm$ 20 & 560 $\pm$ 10 & 500 $\pm$ 10 & 520 $\pm$ 30 & 450 $\pm$ 20 \\
 6731 & [SII] & 220 $\pm$ 10 & 240 $\pm$ 20 & 440 $\pm$ 10 & 360 $\pm$ 10 & 370 $\pm$ 20 & 320 $\pm$ 10 \\
 7135 & [ArIII] & 64 $\pm$ 5 & 59 $\pm$ 4 & 42 $\pm$ 8 & 64 $\pm$ 3 & 51 $\pm$ 5 & 67 $\pm$ 5 \\
 9069 & [SIII] & 210 $\pm$ 20 & 200 $\pm$ 20 & 300 $\pm$ 20 & 180 $\pm$ 10 & 210 $\pm$ 20 & 230 $\pm$ 20 \\
 9229 &  P9 & 26 $\pm$ 3 & 27 $\pm$ 3 & 25 $\pm$ 3 & -- & -- & 27 $\pm$ 3   \\
 9532 & [SIII] & 590 $\pm$ 40 & 500 $\pm$ 60 & 750 $\pm$ 50 & 450 $\pm$ 50 & 470 $\pm$ 40 & 540 $\pm$ 50 \\
 c(H$\beta$)&  & 0.35 & 0.39 & 0: & 0.7 & 0.18 & 0:   \\
 {F(H$\alpha$)
 \footnote{10$^{-16}$ erg cm$^{-2}$ s$^{-1}$, corrected for reddening}}&  & 189 & 140 & 220 & 395 & 93 & 60  \\
 EW(H$\beta$)({\AA}) & & 52 & 46 & 55 & 21 & 14 & 20
\end{tabular}
\end{minipage}
\end{table*}



\subsection{Physical conditions of the gas}

Electron densities for each observed region were derived from the ratio
of the [S{\sevensize II}] $\lambda\lambda$ 6717, 6731 lines following standard 
methods (see for example Osterbrock 1989). For regions GA2 and 74C  
densities of 160 cm$^{-3}$ and 60 cm$^{-3}$ respectively are 
obtained. For the rest of the regions the ratio of the [S{\sevensize II}]
lines implies n$_{e}$ $\leq$ 40 cm$^{-3}$ . 

For two of the observed regions, 74C and 69C, it has been possible to measure the weak auroral [S{\sevensize III}] $\lambda$ 6312 {\AA} line, which, together with the observed intensities of the nebular lines at $\lambda\lambda$ 9069, 9532 {\AA}, can be used to obtain a value for the electron temperature. 
Using the expressions given by Osterbrock (1989) and the effective collision
strengths  recently calculated by Tayal (1997) we find for t(S$^{\rm 2+}$)
values of 7500 $\pm$ 300 K and 7600 $\pm$ 500 K for regions 74C and 69C 
respectively.

Photoionization models indicate that, for abundances close to solar,  the region where sulphur is twice ionized  overlaps both the regions of once and 
twice ionized oxygen and therefore t(S$^{\rm 2+}$) is expected to be 
intermediate between t(O$^+$) and t(O$^{\rm 2+}$). Garnett (1992) from single star 
photoionization models finds a linear relation between the temperatures of O$^{\rm 2+}$ and S$^{\rm 2+}$
\begin{equation}
t(O^{\rm 2+}) = 1.20 t(S^{\rm 2+}) - 0.20 
\end{equation}
We have used that relation to obtain t($O^{\rm 2+}$). Finally, Stasi\'nska's 
(1980) relation between t(O$^{\rm 2+}$) and t(O$^{+}$) has been used to 
calculate the temperature of the t(O$^{+}$) zone. These different temperatures
for regions 74C and 69C are listed in Table 4.  

Another independent way of temperature determination is through the measure of the 
Paschen discontinuity at $\lambda$ 8200 \AA . The ratio of the Paschen discontinuity to the H$\beta$ intensity is, after correcting for the derived extinction value of c(H$\beta$) = 0.48,
\begin{equation}
\frac{\Delta Pa}{I(H\beta)} = (6.9 \pm 0.8) \times 10^{-15} Hz^{-1}
\end{equation} 
 According to the models computed 
by Gonz\'alez-Delgado et al. \shortcite{gode}, for the helium abundance derived for the region 74C (see next section), this 
$\Delta$ Pa/I(H$\beta$) implies an electron temperature of 8000 $\pm$ 700 K,
which agrees with that derived from the [S{\sevensize III}] lines within the errors. Therefore, no temperature fluctuations are apparent in the region which 
might have been expected from the presence of WR stars (P\'erez 1997). 

For the rest of the regions it was not possible to obtain a direct measure of 
the electron temperature. An average temperature has been adopted from an 
empirical calibration by means of the sulphur abundance parameter S$_{\rm 23}$
(S$_{23}$ = ([S{\sevensize II}]6717, 6731 + [S{\sevensize III}]9069, 9532)/H$\beta$)
 (D\'\i az \& P\'erez-Montero 1999). The values for these temperatures, together with similar ones derived for regions 74C and 69C, are also given in Table 4.

\subsection{Chemical abundances}

Ionic abundances of oxygen, nitrogen and sulphur  have been derived following standard methods \cite{pa} and using the temperatures 
found above. These abundances are  listed in Table 4. We have assumed that most 
of the oxygen and sulphur are in the first and second ionization stages and 
therefore O/H = O$^+$/H$^+$ + O$^{++}$/H$^{+}$ and S/H = S$^+$/H$^+$ + S$^{++}$/H$^{+}$. This assumption seems to be justified given the relatively 
low estimates of the electron temperature. 

The helium abundance was determined from the He{\sevensize I} $\lambda$ 4471 and 6678 {\AA} lines.
Mean values of 0.070 $\pm$ 0.005 and 0.060 $\pm$ 0.005 respectively are obtained for regions 74C and 69C.
We have estimated the contribution of neutral helium from the expression:
\begin{equation}
 He^{\rm o} + \frac{He^+}{H^+} = \left(1-0.25 \frac{O^+}{O}\right) ^{-1} \frac{He^+}{H^+} 
\end{equation}
(Kunth \& Sargent 1983). The correction factors obtained for regions 74C and 69C are 1.17 and 1.20 respectively which yield He abundances
y=0.081 for region 74C and y=0.068 for region 69C.


\begin{table*}
 \begin{minipage}{200mm}
 \caption{ Derived physical conditions of the gas in the observed HII regions}
 \scriptsize{
 \begin{tabular}{lcccccccc}
 Parameter  & 74C & 69C & 5N(A) & 5N(B) & GA1 & GA2 & GA3 & GA4 \\
            &     &     &       &       &     &     &     &     \\
 n$_{e}$    & 60  & $\leq$ 40 & $\leq$ 40 & $\leq$ 40 & $\leq$ 40 & 160 & $\leq$ 40 & $\leq$ 40 \\
 t(S$^{2+}$) & 0.75 $\pm$ 0.03 & 0.76 $\pm$ 0.05 & -- & -- & -- & -- & -- & -- \\
 t(O$^{2+}$) & 0.69 $\pm$ 0.04 & 0.72 $\pm$ 0.05 & -- & -- & -- & -- & -- & -- \\
 t(O$^{+})$  & 0.79 $\pm$ 0.03 & 0.80 $\pm$ 0.04 & -- & -- & -- & -- & -- & -- \\
 $<t>_{adop}$& 0.79 & 0.80 & 0.85 & 0.86 & 0.68 & 0.80 & 0.75 & 0.75 \\   
   12 + log(O$^{2+}$/H$^{+}$) & 8.24 $\pm$ 0.14 & 8.08 $\pm$ 0.14 & 7.74 $\pm$ 0.15 & 7.71 $\pm$ 0.15 & 8.15 $\pm$ 0.15 & 7.85 $\pm$ 0.15 & 8.10 $\pm$ 0.15 & 7.74 $\pm$ 0.15 \\
 12 + log(O$^{+}$/H$^{+}$)  & 8.39 $\pm$ 0.09 & 8.38 $\pm$ 0.14 & 8.30 $\pm$ 0.15 & 8.29 $\pm$ 0.15 & 8.47 $\pm$ 0.15 & 8.35 $\pm$ 0.15 & 8.28 $\pm$ 0.15 & 8.40 $\pm$ 0.15 \\
 12 + log(O/H)              & 8.62 $\pm$ 0.11 & 8.56 $\pm$ 0.15 & 8.41 $\pm$ 0.15 & 8.39 $\pm$ 0.15 & 8.65 $\pm$ 0.15 & 8.47 $\pm$ 0.15 & 8.50 $\pm$ 0.15 & 8.49 $\pm$ 0.15\\
 12 + log(S$^{+}$/H$^{+}$)  & 6.29 $\pm$ 0.06 & 6.40 $\pm$ 0.08 & 6.07 & 6.10 & 6.49 & 6.31 & 6.37 & 6.31 \\
 12 + log(S$^{2+}$/H$^{+}$) & 6.75 $\pm$ 0.05 & 6.58 $\pm$ 0.10 & 6.56 & 6.50 & 6.83 & 6.52 & 6.61 & 6.66 \\
 12 + log(S/H)              & 6.88 $\pm$ 0.05 & 6.80 $\pm$ 0.09 & 6.68 & 6.65 & 6.99 & 6.73 & 6.81 & 6.82 \\
 log(N/O)                   &-1.03 $\pm$ 0.06 &-0.94 $\pm$ 0.08 &-1.01 &-1.02 &-0.87 &-0.89 &-0.79 &-0.94 \\
 log(S/O)                   &-1.74 $\pm$ 0.14 &-1.76 $\pm$ 0.20 &-1.73 &-1.74 &-1.66 &-1.74 &-1.69 &-1.67
\end{tabular}
}
\end{minipage}
\end{table*}

\subsection{Wolf-Rayet features}

A prominent Wolf-Rayet feature has been observed in region 74C (see Figure 6).
Assuming a distance to NGC~4258 of 5.5 Mpc \cite{mar} and a constant 
extinction value through this region of c(H$\beta$) = 0.48, as derived from 
the Balmer and Paschen recombination lines, the total luminosity of the broad 
He{\sevensize II} feature, without the contribution of the [Fe{\sevensize III}]
$\lambda$ 4658 {\AA} line, is (1.1 $\pm$ 0.1) $\times$ 10$^{38}$ erg s$^{-1}$. 
This value comprises the broad features of both N{\sevensize III} 
$\lambda\lambda$ 4634, 4640 and He{\sevensize II} $\lambda$ 4686 {\AA} lines. 
The contribution of the N{\sevensize III} lines to the WR bump is 
metallicity-dependent according to Smith \shortcite{smith}. In our case this 
contribution represents 0.4 times the total emission; hence we obtain for the 
He{\sevensize II} line luminosity:

\begin{equation}
L(He{\sevensize II} \lambda 4686) = (6.6 \pm 0.6) \times 10^{37} erg ~ s^{-1} 
\end{equation}

The non-detection of N{\sevensize V} $\lambda\lambda$ 4604,4620 {\AA}, suggests
that the 
WR stars are of late N or intermediate type. Using the calibration of Vacca8\&
Conti \shortcite{vacco}, 38 $\pm$ 5 WN stars are found in this region.


\begin{figure}
 \psfig{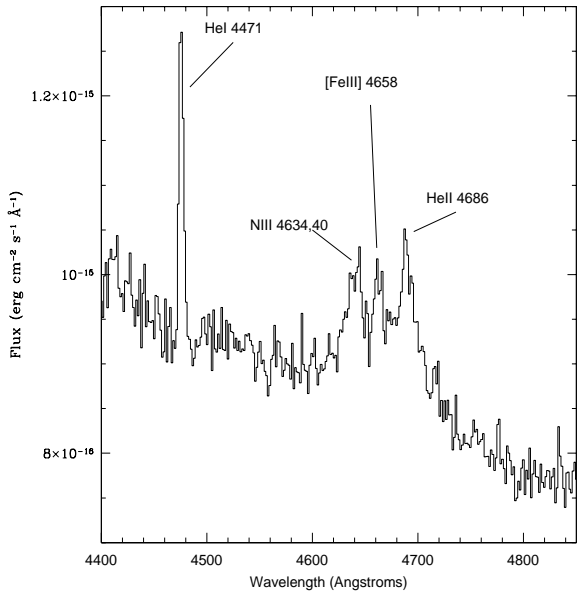}
 \caption{Wolf-Rayet feature at $\lambda$ 4686 {\AA} in region 74C}
\end{figure}


Another fainter Wolf-Rayet feature has been observed in region 5NA. The bump 
luminosity is 2.2 $\times$ 10$^{36}$ erg s$^{-1}$. The N{\sevensize III} lines
contribution represent 0.45 times the total emission, hence the He{\sevensize 
II} line luminosity is 1.2 $\times$ 10$^{36}$ erg s$^{-1}$, which is compatible
with the presence of 1 WN star.

\section{Functional parameters of the observed HII regions}

Three are the fundamental parameters which control the emission line spectra 
of H{\sevensize II} regions \cite{ditevi}:  the ionization parameter, the shape
 of the ionizing continuum, and the metallicity.

The ionization parameter -- {\it i. e.} the ratio of the ionizing photon density to the particle density -- is a measure of 
the degree of ionization of the nebula and can be deduced from the ratio of two lines of the same element corresponding to two different 
ionization states,  e.g. [O{\sevensize II}]/[O{\sevensize III}] or 
[S{\sevensize II}]/[S{\sevensize III}]. Alternatively, it can also be 
determined from [O{\sevensize II}/H$\beta$] or [S{\sevensize II}/H$\alpha$] if the metallicity of the region is known (D\'\i az 1994). By using a large 
grid of single star photoionization models the following expressions are found
(D\'\i az et al. 1991, D\'\i az 1999):
\begin{equation}
 logU = -1.39log([O{\sevensize II}]/H\beta)+0.87log(Z/Z_{\odot})-1.68 
\end{equation}
\begin{equation}
logU = -1.40log([S{\sevensize II}]/H\beta)+1.10log(Z/Z_{\odot})-3.26 \end{equation}
\begin{equation}
logU = -0.80log([O{\sevensize II}]/[O{\sevensize III]}) - 3.02 
\end{equation}
\begin{equation}
logU = -1.68log([S{\sevensize II}]/[S{\sevensize III]}) - 2.99 
\end{equation}

We have derived $U$ from the four expressions above. In all cases the value of U derived from the [O{\sevensize II}]/[O{\sevensize III}] ratio is systematically 
lower than the rest thus implying low effective temperatures for the ionizing 
stars (see {\it e.g.} D\'\i az 1999). We have therefore discarded this value 
and computed U as the mean of the other three. These adopted ionization 
parameters -- $logU$ -- are listed in Table 5 and their uncertainty is estimated to be around $\pm$ 0.2 dex.

 The shape of the ionizing continuum is directly related to the effective temperature of the stars that dominate the radiation field responsible for the 
ionization of the nebula.
A recent version of the photoionization code CLOUDY \cite{fer} has been used to estimate the mean effective temperature of these stars. We have used Mihalas NLTE single-star stellar atmosphere models, with a closed geometry and a constant particle density through the nebula. For all the regions, consistency is found for effective temperatures around 35,000 K - 36,000 K. The results from these photoionization models compared to the observations are presented in Table 5.

\begin{table*}
\vbox to220mm{\vfil Landscape table 5 to go here.
\caption{}
\vfil}
\end{table*}


\section{Discussion}


The H$\alpha$ fluxes for regions 74C and 69C, uncorrected for reddening, are 
2692 and 287 $\times$ 10$^{-16}$ erg s$^{-1}$ cm$^{-2}$ respectively. These 
fluxes are to be compared with 3492 and 332 $\times$ 10$^{-16}$ erg s$^{-1}$ 
cm$^{-2}$ measured by C93. Considering that our observations have been obtained
through a narrow slit and region 74C is rather extended, the agreement can be 
considered satisfactory. The combined H$\alpha$ flux of our regions 5NA and 
5NB is 187 $\times$ 10$^{-16}$ erg s$^{-1}$ cm$^{-2}$ close to the value of 
249 $\times$ 10$^{-16}$ erg s$^{-1}$ cm$^{-2}$ given by C93 for region 5N. 
Regarding the rest of the regions, we tentatively identify regions GA2, GA3 
and GA4 with regions 59C, 58C, 54C of C93. Region GA1 probably corresponds to 
the outer parts of region 69C. 

For each of our observed regions we have calculated the H$\alpha$ luminosity 
and the number of hydrogen ionizing photons from the observed H$\alpha$ flux 
corrected for their derived reddening (see Tables 2 and 3). Both quantities 
depend on the distance D to NGC~4258 according to the expressions:
\begin{equation}
L(H\alpha) = 3.62 \times 10^{37} \left( \frac{F(H\alpha)}{10^{-14}} \right)
\left( \frac{D}{5.5}\right) ^{2} erg \, s^{-1} 
\end{equation}
\begin{equation}
Q(H) = 2.65 \times 10^{49} \left( \frac{F(H\alpha)}{10^{-14}} \right)
\left( \frac{D}{5.5} \right) ^{2} photons \, s^{-1} 
\end{equation}
where F(H$\alpha$) is expressed in erg s$^{-1}$ cm$^{-2}$ and D is given in 
Mpc.

These values are given in Table 6 for the adopted distance of 5.5 Kpc. Only 
region 74C has an H$\alpha$ luminosity greater than 10$^{39}$ erg s$^{-1}$ and 
can be classified as a supergiant HII region as defined by Kennicutt (1983). 
The rest of the regions have H$\alpha$ luminosities typical of HII regions 
in early spiral galaxies, although all of them are greater than 10$^{37}$ 
erg s$^{-1}$ (Q(H) $>$ 10$^{49}$ photons s$^{-1}$) requiring more than a single
star for their ionization (Panagia 1973).


Filling factors for each observed region can be determined from the reddening 
corrected  H$\alpha$ flux, F(H$\alpha$), and the derived ionization parameter, 
U,  according to the expression:

\begin{eqnarray}
 \epsilon =& 0.34 \left(\frac{F(H\alpha)}{10^{-14}}\right)^{-1/2}
\left(\frac{D}{5.5}\right)^{-1} \left(\frac{U}{10^{-3}}\right)^{3/2} \\
& \left(\frac{\alpha _B(H^o,T)}{10^{-13}}\right)^{-1} \left(\frac{n_e}{100}\right)^{-1/2}\nonumber
\end{eqnarray}
where $\alpha _B(H^o,T)$ is the recombination coefficient for hydrogen and 
$n_e$ is the electron density of the emitting gas. The fillin0 factors 
computed in this way are listed in Table 6. We have used a value of 
$\alpha _B(H^o,T)$ = 3.76 $\times$ 10$^{-13}$ cm$^{3}$ s$^{-1}$, corresponding 
to T= 7000 K and n$_e$= 100 cm$^{-3}$ (Osterbrock 1989). For regions 74C and 
GA2 we have used values of n$_e$= 60 and 160 cm$^{-3}$ respectively, as 
derived from the ratio between the red [SII] lines. For the rest of the regions
for which only upper limits to the density could be derived, a value of 
n$_e$=10 cm$^{-3}$ has been assumed.


\begin{table*}
 \begin{minipage}{120mm}
 \caption{ Physical properties of the observed HII regions}
 \begin{tabular}{lccccc}
Region & L(H$\alpha$) & Q(H) & $\epsilon$ & M(HII) & M$^{\star}$ \\
       & (10$^{38}$ erg s$^{-1}$) & (10$^{49}$ s$^{-1}$) & & (M$_{\odot}$) & (M$_{\odot}$) \\
74C     &  32.1   & 148     & 0.10  & 78200 & 112000 \\
69C     &   1.85  &   8.49  & 0.45  & 27000 &   4840 \\
5NA     &   1.09  &   5.01  & 0.41  & 15900 &   5570 \\
5NB     &   0.81  &   3.70  & 0.34  & 11800 &   4570 \\
GA1     &   1.27  &   5.84  & 0.27  & 18500 &   6180 \\
GA2     &   2.27  &  10.4   & 0.04  &  2080 &  25200 \\
GA3     &   0.54  &   2.46  & 0.29  &  7830 &   8450 \\
GA4     &   0.35  &   1.59  & 0.52  &  5050 &   4010 \\      
 \end{tabular}
 \end{minipage}
\end{table*}


According to C93 the 
angular sizes of regions 74C, 69C and 5N are 38.6, 16.8 and 20.5 arcsec 
respectively. The corresponding linear sizes, at the assumed distance to 
NGC~4258, are 1.0, 0.45 and 0.55 Kpc . These are comparable to the sizes of 
\HII\ regions of moderate intensity found in other spiral galaxies 
(e.g.~Gonz\'alez-Delgado et al. 1997).

H$\alpha$ angular effective diameters -- that is the diameters containing half 
the H$\alpha$ emission -- of regions 74C, 69C and 5N are 4.3, 3.7 and 6.1 
arcsec respectively (C93) which translates into 115, 99 and 163 pc.
From the definition of U it is possible to obtain the angular size of the 
emitting region using the reddening corrected H$\alpha$ flux and the derived 
electron density:
\begin{equation}
\left( \frac{\phi}{10^{\prime\prime}}\right) = 0.06 \left(\frac{F(H\alpha)}
{10^{-14}}\right)^{1/2} \left(\frac{U}{10^{-3}}\right)^{-1/2} \left(\frac{n_e}
{100}\right)^{-1/2} 
\end{equation}
where $\phi$ is the angular diameter in units of 10 arcsec . This angular 
diameter does not depend on the assumed distance to the galaxy. For region 
74C the derived angular diameter is 3.3 $^{\prime\prime}$  close to the value 
of 4.3 $^{\prime\prime}$ given by C93. For region 69C their observed value of 
the effective diameter of 3.7 $^{\prime\prime}$ yields a value of the electron 
density of n$_e$ = 5 cm$^{-3}$ close to the assumed one of 10 cm$^{-3}$. Using 
this value of n$_e$ for the rest of the observed regions, except GA2 for which 
n$_e$ = 160 cm$^{-3}$ we obtain angular diameters between 1 and 2.6 arcsec.


We can also calculate the mass of ionized hydrogen  following the 
expression: 
\begin{equation}
 M(\HII) = \frac{m_p Q(H)}{n_e (1+y^+) \alpha _B(H^o,T)} 
\end{equation}
(Osterbrock 1989) where m$_p$ is the mass of the proton and $y$ the abundance 
by number of ionized helium, that we have taken as 0.10. Using observed and 
derived quantities this expression becomes 
\begin{equation}
 M(HII) = 538 \left(\frac{F(H\alpha)}{10^{-14}}\right) \left(\frac{n_e}{100}\right)^{-1} \left(\frac{D}{5.5}\right)^{2} M_{\odot} 
\end{equation}
The corresponding masses of ionized hydrogen in each observed region are 
given in Table 6.


From evolutionary models of single ionizing clusters and radiation bounded 
H{\sevensize II} regions, a relation between the number of ionizing Ly$\alpha$ 
photons per second per solar mass and the H$\beta$ equivalent width 
\cite{diaz} can be found, which allows the estimation of the ionizing cluster 
mass by taking into account the cluster evolution.
\begin{equation}
log(Q(H)/M_{\odot}) = 0.86 log(EW(H\beta)) + 44.48 
\end{equation}
 For our observed regions all values for Q(H) are around 10$^{49}$ photon s$^{-1}$ except for region 74C for which a value of the order of 10$^{51}$ photon s$^{-1}$ is found. Hence, in the absence of dust, a lower limit for the mass of the ionizing clusters can be estimated by means of the H$\beta$ measured equivalent width and the H$\alpha$ luminosity for each region. The estimated ionizing cluster masses range from 4 $\times$ 10$^{3}$ M$_{\odot}$ for region GA4 to 1.12 $\times$ 10$^{5}$ M$_{\odot}$ for region 74C.


\begin{figure}
 \psfig{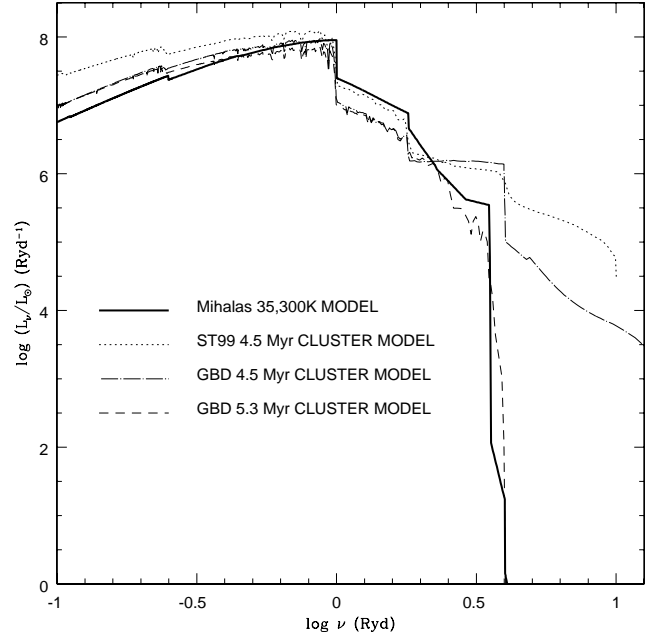}
 \caption{Spectral energy distribution of different ionizing clusters for 
region 74C}
\end{figure}



The high signal-to-noise spectrum of this latter region allows a more detailed 
modelling. The presence of WR features, in principle, provides a means to 
constrain the age of the ionizing population. The observed WR ``bump" 
luminosity relative to H$\beta$ (L(WR)/H$\beta$) is 0.15. According to the 
models by Schaerer \& Vacca (1998), at metallicity Z= 0.008 (0.4 solar), these 
high values are only found at an age of 4.5 Myr. However, this ratio is 
strongly dependent on metallicity and changes by more than an
order of magnitude between Z=0.008 (0.4 solar) and Z= 0.02 (solar). Therefore, 
the slightly higher metallicity of region 74C (Z $\simeq$ 0.012) might allow 
earlier ages for the WR stellar population. Clearly, the available grid for 
stellar evolution models is too coarse to elucidate this matter. The same 
applies to the equivalent width of the ``bump" (EW(WR)) whose observed value 
is 10.8 \AA . Since the NIII $\lambda$ 4640 \AA\ line is strongly dependent on 
metallicity (Smith et al. 1996) it is probably more advisable to use only 
the flux and equivalent width of the HeII line. In that case, our observed 
values of L(HeII)/H$\beta$ and EW(HeII) are  0.09 and 6.6 \AA\ respectively.
Again, consistency is found for a model of Z=0.008 and an age of 4.5 Myr.

 Since, according to the recent evolutionary models of Leitherer et al. (1999; 
STARBURST99), the observed value of the equivalent width of H$\beta$ (81 \AA ) 
points also to an age of around 4.5 Myr, we have computed the emission 
line spectrum corresponding to this single ionizing cluster.
\begin{footnote}{ All the models are computed assuming a standard Salpeter IMF for the stellar ionizing cluster and using the photoionization code CLOUDY (Ferland 1996).} 
\end{footnote}
This set of models makes use of the same evolutionary tracks as the 
models of Schaerer \& Vacca (1998) and therefore provides a selfconsistent  
frame. The results of the computation can be seen in Table 7 together with 
the observed emission line ratios. It can be seen that the model ionizing 
spectrum results too hard and does not reproduce the observations. This is not 
surprising in view of the low effective temperature found from single star 
photoionisation models (Table 5). It can be better appreciated in Fig 8 where 
the spectral energy distribution of the cluster can be compared to that of the 
Mihalas stellar atmosphere of a star with T$_{eff}$ = 35300 K which adequately 
reproduces the data. A high energy tail is clearly present in the ionising cluster as the result of the WR star contribution.

%
%

\begin{table*}
\setcounter{table}{6}
 \begin{minipage}{150mm}
  \caption{Evolutionary models for region 74C}
  \begin{tabular}{@{}lccc@{}}
   Parameter & Observations & 5.3 Myr Model (GBD) & 4.5 Myr Model (ST99) \\
 3727 [OII] & 2100 $\pm$ 40 & 2396 & 2181 \\
 5007 [OIII] & 1140 $\pm$ 10 & 1366 & 2394 \\
 4959 [OIII] & 378 $\pm$ 5 & 473 & 829 \\
 6548 [NII] & 200 $\pm$ 10 & 200 & 242 \\ 
 6584 [NII] & 610 $\pm$ 20 & 600 & 713 \\
 6717 [SII] & 260 $\pm$ 10 & 217 & 386 \\
 6731 [SII] & 188 $\pm$ 3 & 160 & 282 \\
 6312 [SIII] & 9 $\pm$ 1 & 9 & 8 \\
 9069 [SIII] & 266 $\pm$ 4 & 333 & 308 \\
 9532 [SIII] & 660 $\pm$ 20 & 827 & 765 \\ 
 EW(H$\beta$)({\AA}) & 81 & 100 & 97 \\
 log Q(H) & 51.17 & 51.17 & 51.17 \\
 $<$log U$>$ & -2.45 $\pm$ 0.05 & -2.30 & -2.48 \\
 t(O$^{+}$) & 0.79 $\pm$ 0.03 & 0.76 & 0.81 \\
 t(O$^{2+}$) & 0.69 $\pm$ 0.04 & 0.66 & 0.69 \\
 Z/Z$_{\odot}$ & 0.5 & 0.65 & 0.65
\end{tabular}
\end{minipage}
\end{table*}


These models assume an enhanced mass loss during the evolution of massive stars, as prescribed by the stellar evolution models of Meynet et al. (1994). More 
moderate mass loss rates are assumed in the models of Bressan et al. (1993) 
and Fagotto et al. (1994) and with them a single population of 5.3 Myr 
can adequately reproduce the emission line spectrum (see Table 7). 
The corresponding ionizing spectrum is also shown in Fig 8. However, this ionizing cluster provides only a few WR stars (about 3$\times$ 10$^{-5}$ per 
solar mass of the ionizing cluster, for a metallicity Z=0.008; Garc\'\i a-Vargas, 
Bressan \& D\'\i az 1995). This is 
probably not enough to reproduce the observed values of L(WR)/H$\beta$.

The WR features observed in region 74C are similar to 
those found in other extragalactic HII regions. Table 8 shows the values of 
the WR feature intensities and equivalent widths for 74C and other three well 
studied extragalactic HII regions: NGC~604 in M33 (D\'\i az et al 1987; Terlevich et al. 1996),
region A in NGC~3310 (Pastoriza et al. 1993) and region A in NGC~7714 
(Garc\'\i a-Vargas et al. 1997). Also given in the table are the metallicities 
of the regions in the form 12+log(O/H) and the effective temperature of the 
ionizing radiation estimated from single star photo-ionization models.


\begin{table*}
 \begin{minipage}{150mm}
 \caption{ WR feature intensities and equivalent widths in region 74C and                   other extragalactic \HII\ regions.}
 \begin{tabular}{lcccccc}
Region & L(WR)/H$\beta$ & EW(WR)(\AA ) & L(HeII)/H$\beta$ & EW(HeII)(\AA ) & 
12+log(O/H) & T$_{eff}$ (K) \\

Region A (NGC~3310)   & 0.16 &  5.7 & 0.11 & 3.9 & 8.21 & 40000 \\ 
NGC~604 (M33)         & 0.12 & 10.3 & 0.06 & 5.2 & 8.40 & 37000 \\
Region A (NGC~7714)   & 0.07 &  1.5 & 0.05 & 1.0 & 8.47 & 36000 \\
Region 74C (NGC~4258) & 0.15 & 10.8 & 0.09 & 6.6 & 8.62 & 35300 \\

 \end{tabular}
 \end{minipage}
 \end{table*}

The tabulated values are difficult to explain in the light of the recent models 
of Schaerer \& Vacca (1998). The expected decrease of the intensities of WR 
features with decreasing metallicity is not clearly observed. 
However, a decrease of the 
temperature of the ionizing radiation with increasing metallicity is apparent.  
In fact, region A in NGC~3310 and region 74C in NGC~4258 show the same L(WR)/H$\beta$ ratio despite their 
different metallicity (by a factor of about 3). This implies that both regions 
have similar ratio of WR to O stars, while, at the same time, their effective 
temperatures are considerably different.

The interpretation of the equivalent widths of WR features is even more difficult, since they may be affected by the continuum from any underlying stellar population. Composite populations have been postulated for region A in 
NGC~3310 (Pastoriza et al 1993) and region A in NGC~7714 (Garc\'\i a-Vargas et al. 1997). Under this assumption,  the equivalent widths, EW(WR) and EW(HeII), of the cluster
with WR stars are increased to 6  and 4.8 \AA\ for region A in NGC~3310 
and 3.6 and 2.5 \AA\ for region A in NGC~7714 which might still be compatible with Scharer \& Vacca models. In the case of region 74C, 
combinations of two ionizing clusters can be found which adequately reproduce 
the emission line spectrum; however, this assumption would produce equivalent 
widths for the WR cluster higher than the maximum predicted for the models with Z=0.008, although probably compatible with solar metallicity models. Again, a finer metallicity grid for the WR cluster synthesis models would be needed in 
order to explore this possibility in further detail. 

\section{Summary and conclusions} 

We have analyzed eight H{\sevensize II} regions in the LINER galaxy NGC~4258 using spectrophotometric observations between 3700 and 9650 {\AA}. For two of the regions it has been possible to measure the electron temperature from the [S{\sevensize III}] $\lambda$ 6312 {\AA} line, which allows the derivation of accurate abundances following standard methods. For the rest of the regions an empirical calibration based on the 
sulphur emission lines has been used to determine a mean oxygen content. 
The derived metallicities range from 0.3 to 0.6 Z$_{\odot}$. In particular, 
the metallicities found for regions 74C and 5N, previously reported by Oey \& 
Kennicutt (1993) to be close to solar, are found to be lower by a factor 
of two.

For each observed region, we have also estimated the functional parameters: ionization parameter and the effective temperature of the ionizing cluster. 
Most regions show ionization parameters of the order of 10$^{-3}$ and 
effective temperatures of around 35300 K, except regions 74C and 69C for which 
higher ionization parameters are found. We have also derived the physical 
properties of the regions and their corresponding ionizing clusters: fillin0 factor, mass of ionized gas and mass of ionizing stars. Most of the regions, except 74C, have small ionizing clusters with masses in the range 4000 to 
25000 solar masses. These values constitute in fact lower limits since the 
regions are assumed to be ionization bounded and the presence of dust has not been taken into account.

WR features have been detected in region 74C and, to a lower extent, in region 
5NA. 
A detailed modelling has been carried out for region 74C using different sets 
of models: those of Schaerer \& Vacca (1998) for WR populations and those of 
Leitherer et al. (1999) and Garc\'\i a-Vargas, Bressan \& D\'\i az (1995) for 
ionizing populations, to try to reproduce simultaneously both the WR features 
and the emission line spectrum. No consistent solution has been found: the 
models which better reproduce the WR features (a cluster 4.5 Myr old in the 
models of Schaerer \& Vacc8 and Leitherer et al.) result too hard to reproduce 
the emission line spectrum. Conversely, a cluster 5.3 Myr old in the models of 
Garc\'\i a-Vargas, Bressan \& D\'\i az (1995) adequately reproduces the 
emission line spectrum, but produces too few WR stars to explain the observed 
WR features.

When comparing the WR ``bump" and HeII intensities and equivalent widths with 
other well studied HII regions of different metallicities, the data on region 74C are found to fall into the observed ranges. However, the expected 
decrease of WR feauture intensities with decreasing metallicity is not clearly 
observed. On the other hand a decrease in the effective temperature of the ionizing 
radiation with increasing metallicity is apparent. This seems to imply that, 
while the number ratio of WR to O stars is similar for regions of metallicities 
differing by a factor of 3, the WR stars are cooler for the regions with higher 
metallicity.

Both more observations of confirmed high metallicity regions and a finer 
metallicity grid for the evolutionary synthesis models are needed in order to 
understand the ionizing populations of HII regions.

\section*{Acknowledgements}

The WHT is operated in the island of La Palma by the Issac Newton Group in the Spanish Observatorio del 
Roque de los Muchachos of the Instituto de Astrof\'\i sica de Canarias. We 
would like to thank CAT for awarding observing time. We also thank an anonymous 
referee for helpful suggestions.

E.T. is grateful to an IBERDROLA Visiting Professorship 
to UAM during which part of this work was completed.
This work has been partially supported by DGICYT project PB-96-052.



\end{document}



-- Escribir parrafo sobre la posibilidad de usar los modelos de Cervinio y Miguel Mas.

-- Referencia a poblaciones compuestas en otras regiones HII observadas.

-- Partir la discusion?

-- Conclusiones.

__ Cambiar abstract?

\subsection{Mass of the ionizing clusters}
Filling factors have been derived from the dimensionless ionization parameter U and the total number of recombinations Q(H). Hence the fillin0 factor is defined as\\
$\epsilon^{2/3}$ = (4$\pi$/Q(H)n)$^{1/3}$ $\times$ (3/$\alpha_{B}$(H$^{0}$))$^{2/3}$ $\times$ cU,  where $\alpha_{B}$(H$^{0}$) is the recombination coefficient for hydrogen, c is the speed of light and n is the particle density. For an electron temperature of 8,500K, the filling factors obtained are 0.052 and 0.31 for regions 74C and 69C, respectively. Another way to determine $\epsilon$ is by means of the algorythm derived by D\'\i az et al. \shortcite{ditevi} as a function of the angular diameter of the regions, the distance to the galaxy, the H$\beta$ intensity and the ionization parameter. Derived values are 0.047 and 0.28 for these two regions in good agreement with the previous ones. These values are common in other H{\sevensize II} regions \cite{ost}. As for the particle density \cite{ditevi}, values of 52 and 7 particles cm$^{-3}$are found in regions 74C and 69C respectively, in excellent agreement with the derived ones from the [S{\sevensize II}] emission lines.


using the algoritm He$^{+}$/H$^{+}$ = 2.04$t^{0.13}$(4471/H$\beta$) where $t$ is the electron temperature in units of 10$^{4}$K.

Hence the low helium abundance could be due to the presence of an underlying stellar population when the He{\sevensize I} 4471 equivalent width is quite small. The He$^{+}$/H${^+}$ abundance ratio can be deduced from the intensities
of the He{\sevensize I} lines. A value of 0.070 $\pm$ 0.006 is obtained (this value will be discussed in the next section). These temperature fluctuations could be due to variations in the local heating and cooling rates because of, for instance, the presence of Wolf-Rayet stars in the region \cite{per}. In our case, a prominent Wolf-Rayet feature has been observed in the region 74C.  We can
conclude that, in spite of the presence of WR stars in the region 74C, no
appreciable temperature fluctuations exist in this region. Anyway, better resolution observations must be done in order to test this result. No measurement of the Paschen discontinuity was possible in the other regions because of the
low S/N ratio.\\

For the rest of the regions in which no direct determination of the
temperature was possible, the oxygen abundances have been derived using an
empirical calibration based on the sulphur emission lines \cite{dpm}. This calibration can be fitted by the regression line: 12 + log (O/H) = 1.265 log S$_{23}$ + 8.252 where S$_{23}$ is the sulphur abundance parameter defined as follows: S$_{23}$ = ([S{\sevensize II}]6717, 6731 + [S{\sevensize III}]9069, 9532)/H$\beta$. The derived abundances and ionization parameters are shown in Table 4. The regions 5N and 74C were previously analyzed by Oey \& Kennicutt \shortcite{ok}. Their derived oxygen abundances by using empirical calibrations based on the optical oxygen lines are 8.74 and 8.87 respectively. These abundances are significantly higher than the ones derived by us. All the observed regions have metallicities that range from 0.2 (74C and 69C) to 0.5 (GA1) solar metallicity. Similar metallicity trends are found in NGC 3310 \cite{PDT}, in which the observed circumnuclear and disk H{\sevensize II} regions have a moderately low metallicity (0.2-0.4 Z$_{\odot}$)and in the GEHRs of NGC 7714 (0.2-0.35 Z$_{\odot}$)\cite{ggp}. From these results, the adopted average oxygen abundance for NGC 4258 (8.97 at $\rho$ = 0.4$\rho_{0}$) \cite{zkh}, must be revised.\\
The N/O and S/O ratios are higher than solar in regions 74C and 69C. The derived N/O ratio is very similar to the one deduced from models for the high metallicity H{\sevensize II} regions in M51 \cite{ditevi}. In our case, we should expect that a secondary production of nitrogen is increased with age. In the case of the S/O ratio, we can conclude that, from our results, a low O/H implies an increase in S/O.  

Hence, an average ionization parameter has been calculated. The mean value is in all cases lower than -2.5. The highest value corresponds to the high excitation region 74C (-2.7). In the other regions, the average value is around -3.0, hence  the ionization parameter is constrained by their relatively low number of ionizing photons which is evident from the H$\alpha$ images of the galaxy. In the case of the region 74C, the ionizing photon density is higher but it is balanced by the electron density (60 cm$^{-3}$).

 However, our metallicity value for this region is Z = 0.004 (0.2 Z$_{\odot}$), hence there would be two possible explanations for this shift in the metallicity: either the models underpredict the WR/O star number ratios for an instantaneous burst and the WR bump intensities relative to H$\beta$ as a function of the H$\beta$ equivalent width or our inherent errors in the determination of the metallicity are greater than expected, due, for example, to small errors in T$_{e}$ that propagates exponentially into the abundance. Anyway, there is a great uncertainty in the values predicted by the models, especially in the low metallicity regime. New observations are necessary to
test current WR models and also a refinement in the models predictions at low metallicities.  
 
 In region 74C, it should be expected that the presence of hot, luminous Wolf-Rayet stars, may increase the mean effective temperature of the ionizing cluster. 


\subsection{Chemical abundances}

\begin{table*}
 \begin{minipage}{100mm}
 \caption{Derived parameters in the observed H{\sevensize II}
 regions}
 \begin{tabular}{@{}lcccccc@{}}
 {Parameter\footnote{All the abundaces given as 12 + log(O/H), except He/H, which is given by number}} & O/H & S/H & N/H & Ne/H & He/H & $<$log U$>$ \\
     &           &     &     &     &    &   \\
 74C & 8.17 $\pm$ 0.09 & 6.72 $\pm$ 0.06 & 7.3 $\pm$ 0.2 & 7.0 $\pm$ 0.2 & 0.066 $\pm$ 0.005 & -2.7 $\pm$ 0.1 \\
 69C & 8.10 $\pm$ 0.1 & 6.62 $\pm$ 0.10 & 7.4 $\pm$ 0.2 & ... & 0.060 $\pm$ 0.005 & -2.9 $\pm$ 0.1 \\
 5N(A) & 8.4 $\pm$ 0.1 & ... & ... & ... & ... & -2.85 $\pm$ 0.05 \\
 5N(B) & 8.4 $\pm$ 0.1 & ... & ... & ... & ... & -2.9 $\pm$ 0.1 \\
 GA1   & 8.6 $\pm$ 0.1 & ... & ... & ... & ... & -3.0 $\pm$ 0.1 \\
 GA2   & 8.4 $\pm$ 0.1 & ... & ... & ... & ... & -3.1 $\pm$ 0.1 \\
 GA3   & 8.5 $\pm$ 0.1 & ... & ... & ... & ... & -3.1 $\pm$ 0.1 \\
 GA4   & 8.5 $\pm$ 0.1 & ... & ... & ... & ... & -3.0 $\pm$ 0.1
 \end{tabular}
 \end{minipage}
\end{table*}

\begin{table}
 \begin{minipage}{200mm}
  \caption{Photoionization models and observations for P.A.= 133\degr}
  \begin{tabular}{@{}lcccc@{}}
   Region & \multicolumn{2}{c}{74C} & \multicolumn{2}{c}{69C} \\
          & Model & Observed & Model & Observed \\
   Parameter &  &  &  & \\
 $<$log U$>$ & -2.50 & -2.7 $\pm$ 0.1 & -2.80 & -2.9 $\pm$ 0.1 \\
 n$_{e}$(cm$^{-3}$) & 50 & 60 & 10 & $<$50 \\
 T$_{eff}$(K) & 35,000 & ... & 35,500 & ... \\
 12 + log(O/H) & 8.05 & 8.17 $\pm$ 0.09 & 8.00 & 8.1 $\pm$ 0.1 \\
 12 + log(N/H) & 7.20 & 7.3 $\pm$ 0.2 & 7.25 & 7.4 $\pm$ 0.2 \\
 12 + log(S/H) & 6.80 & 6.72 $\pm$ 0.06 & 6.70 & 6.62 $\pm$ 0.10 \\
 12 + log(Ne/H) & 7.25 & 7.0 $\pm$ 0.2 & ... & ... \\
 3727 [OII] & 2.13 & 2.10 $\pm$ 0.04 & 2.26 & 2.28 $\pm$ 0.09 \\
 4959,5007[OIII] & 1.54 & 1.52 $\pm$ 0.02 & 1.23 & 1.22 $\pm$ 0.04 \\
 6548,6584[NII] & 0.80 & 0.81 $\pm$ 0.03 & 1.01 & 1.04 $\pm$ 0.03 \\
 6717,6731[SII] & 0.46 & 0.45 $\pm$ 0.02 & 0.63 & 0.62 $\pm$ 0.02 \\
 9069,9532[SIII] & 1.41 & 0.93 $\pm$ 0.03 & 1.09 & 0.67 $\pm$ 0.04 
\end{tabular}
\end{minipage}
\end{table}

\end{document}